\documentclass[aps,prl,twocolumn,showpacs,superscriptaddress,10pt]{revtex4-1}

\usepackage{graphicx}
\usepackage{amsmath,amssymb,amstext,amsthm}
\usepackage{bbold}
\usepackage{cancel}
\usepackage{color}
\usepackage{verbatim}
\usepackage{booktabs}
\usepackage[caption=false]{subfig}
\usepackage[english]{babel}

\usepackage[utf8x]{inputenc}

\begin{document}

\title{Inversion-symmetry protected chiral hinge states \\ in stacks of doped quantum Hall layers}

\author{Sander H. Kooi}
\affiliation{Institute for Theoretical Physics, Center for Extreme Matter and
Emergent Phenomena, Utrecht University, Princetonplein 5, 3584 CC Utrecht,
Netherlands}

\author{Guido van Miert}
\affiliation{Institute for Theoretical Physics, Center for Extreme Matter and
Emergent Phenomena, Utrecht University, Princetonplein 5, 3584 CC Utrecht,
Netherlands}

\author{Carmine Ortix}
\affiliation{Institute for Theoretical Physics, Center for Extreme Matter and
Emergent Phenomena, Utrecht University, Princetonplein 5, 3584 CC Utrecht,
Netherlands}
\affiliation{Dipartimento di Fisica ``E. R. Caianiello", Universit\'a di Salerno, IT-84084 Fisciano, Italy}

\date{\today}

\begin{abstract}

We prove the existence of higher-order topological insulators with protected
chiral hinge modes in quasi-two-dimensional systems made out of coupled layers
 stacked in an inversion-symmetric manner.
In particular, we show that an external magnetic field drives a stack of
alternating p- and n-doped buckled honeycomb layers into a higher-order
topological phase, characterized by a non-trivial three-dimensional ${\mathbb Z}_2$ invariant.
We identify silicene multilayers as a potential material platform for the
experimental detection of this novel topological insulating phase.

\end{abstract}

\maketitle

\paragraph{Introduction --} 

A free-fermion symmetry protected topological (SPT) insulator is a quantum
state of matter that cannot be adiabatically deformed to a trivial atomic
insulator without either closing the insulating bulk band gap or breaking the
protecting symmetry~\cite{bra17,po17}. 
Its topological nature is reflected in the 
general appearance of gapless boundary modes in one dimension lower
~\cite{has10,qi11}.  
However, when the protecting symmetry is a
crystalline symmetry the gapless boundary modes appear only on surfaces which are left
invariant under the protecting symmetry operation~\cite{fu11,hsi12}. 
Most importantly, 
these gapless boundary modes are ``anomalous": on a single
surface the number of fermion flavors explicitly violates the fermion doubling
theorem~\cite{nie81} or stronger version of it~\cite{fan17}.

Very recently, it has been shown that point-group symmetries can stabilize
insulating states of matter in bulk crystals with conventional gapped
surfaces, but with gapless modes at the hinges connecting two surfaces related
by the protecting crystalline symmetry~\cite{ben17,ben17bis,sch18,sch18bis,gei18,xu17,pet18,ser18,imh17,kha18,eza18,lan17,sit12,son17,fan17,eza18bis,eza18cis}. 
For systems of spinless electrons (negligible spin-orbit coupling) the hinge modes are chiral. 
Hence, they represent  anomalous one-dimensional (1D) modes -- they cannot be encountered in any conventional 1D
atomic chain  -- but now embedded in a three-dimensional crystal. These novel
topological crystalline insulators, which have been dubbed higher-order
topological insulators,  have started to be classified in systems possessing
different crystalline symmetries, including rotational and rotoinversion
symmetries~\cite{ben17,ben17bis,son17,lan17,mie18}.

In inversion-symmetric crystals, higher-order topological insulators can also
exist~\cite{kha18}. However, in this case, inversion symmetry-related surfaces
are connected via two hinges, one of which will host a chiral gapless mode.
This, in turn, gives rise to an additional modulo two ambiguity in the
microscopic hinge location of the chiral modes, reminiscent of the ambiguity
in the Fermi arcs connectivity of Weyl
semimetals~\cite{lau17}.

The aim of this Letter is to show that such an inversion-symmetry protected
higher-order topological insulator can be in principle obtained in stacks of
doped buckled honeycomb layers (e.g. silicene~\cite{kam13}) subject to an external magnetic field. Two factors conspire to render this
possible. First, the quantum Hall states in p- and n-doped honeycomb layers
generally have opposite sign of the Hall conductance and hence
are characterized by opposite Chern numbers $\mathcal{C}$~\cite{net09}. Second, in a
simple AA stacking configuration, buckled honeycomb layers intrinsically break
both the reflection symmetry in the stacking direction and the twofold
rotation about the stacking direction, but still preserve the three-dimensional
bulk inversion symmetry. We first give an intuitive argument for the existence
of an inversion- symmetric higher-order topological insulator in stacks of
Chern insulators with alternating $\mathcal{C}=\pm 1$ integer  invariants, and
show that this insulating phase can be derived from a parent mirror Chern
insulator~\cite{ful16} by adding crystalline symmetry-breaking terms. Then, we introduce a
corresponding minimal tight-binding model consisting of quantum anomalous Hall
layer stacks~\cite{hal88}, and verify its topological nature by computing the
corresponding bulk $\mathbb{Z}_2$  invariant~\cite{mie18}. Finally, we perform
a full Hofstadter~\cite{hof76} calculation in three-dimensions~\cite{ber07}
for buckled honeycomb layers to show the existence of topologically protected chiral hinge states.

\begin{figure}
\includegraphics[width=1\columnwidth]{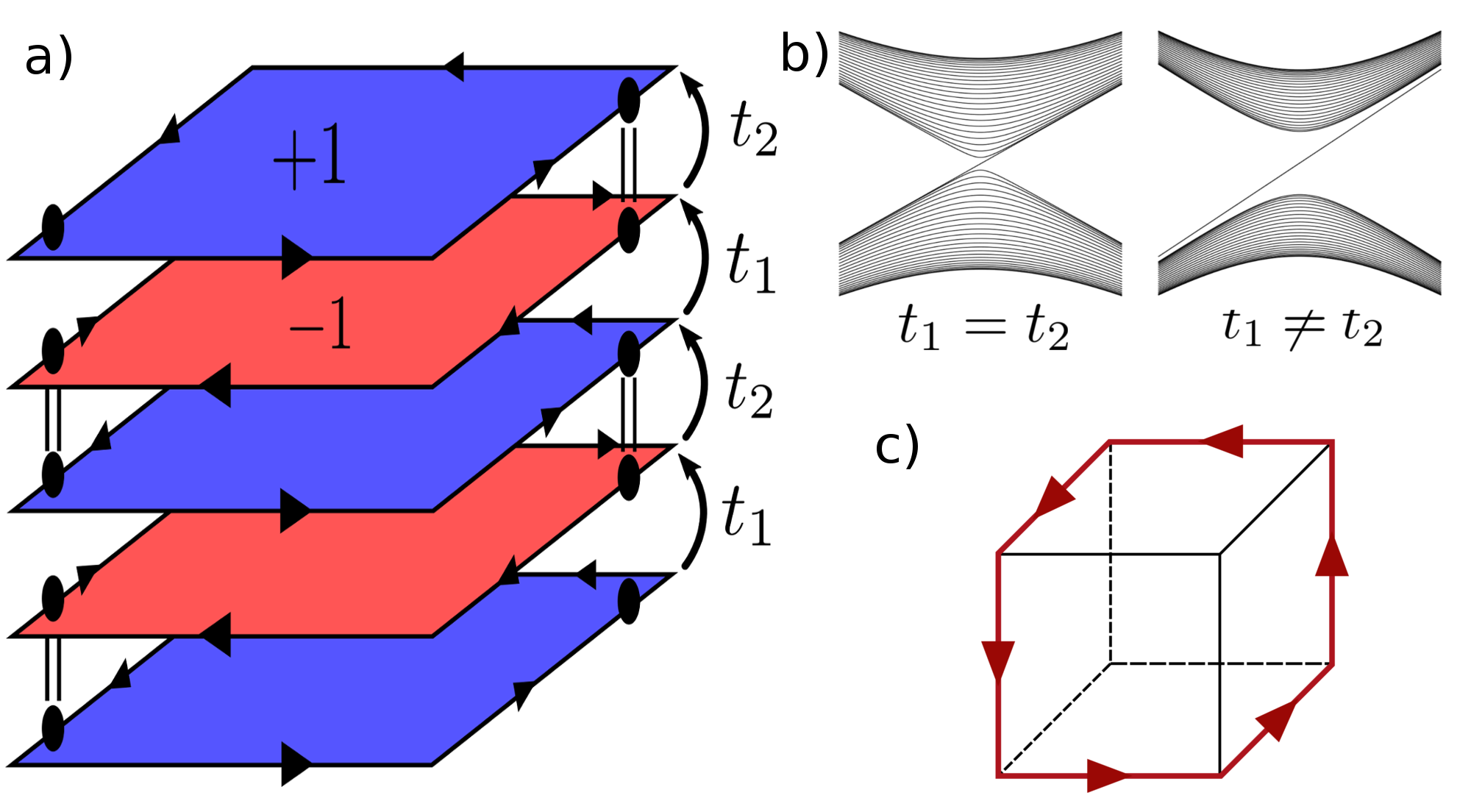}
\caption{a) Sketch of stacked Chern insulators with alternating Chern numbers. We also indicate the effective inversion-symmetric coupling between the chiral edge states. b)
Corresponding surface energy spectrum for mirror symmetric couplings ($t_1=t_2$) and inversion-symmetric couplings ($t_1 \neq t_2$). 
c) Schematic figure of the inversion symmetric hinge states in a cube geometry.
\label{fig:schematic}}
\end{figure}

\paragraph{Effective surface theory --}
Let us start out with an effective low-energy approach for stacks of Chern insulating layers of alternating $\mathcal{C}=\pm 1$ integer topological invariant [c.f. Fig.~\ref{fig:schematic}(a)]. 
At any edge perpendicular to the stacking direction, each Chern insulating
layer is characterized by an anomalous chiral edge mode whose dispersion can
be considered to be linear. For completely uncoupled layers, the effective surface Hamiltonian in the primitive two-layer surface unit cell then reads
${\cal H}_0 = k_x \sigma_z$, where the Pauli matrix acts in the layer space
and we explicitly considered a $(010)$ surface. We next introduce an
interlayer coupling between consecutive layers with a coupling strength,
which, for simplicity, we assume to be real. The effective surface
Hamiltonian is then modified accordingly  to ${\cal H}_{S} = {\cal H}_0 + t
\left[1+\cos{(k_z)}\right] \sigma_x - t \sin{(k_z)} \sigma_y$. It preserves
mirror symmetry in the stacking direction around one layer~\cite{ful16} with the reflection
operator that acquires an explicit momentum dependence and reads
$\mathcal{M}(k_z)=\textrm{diag}\left(1,\mathrm{e}^{-i k_z}\right)$. The mirror
symmetry constraint $\mathcal{M}(k_z) \mathcal{H}_{S} {\mathcal M} (k_z)^{-1}
\equiv \mathcal{H}_{S}\left(k_z \rightarrow -k_z \right)$ implies a  decoupling of the chiral modes on the $k_z= \pi$ line. Hence,
the three-dimensional system is characterized by gapless surfaces with a
mirror-symmetry protected single Dirac cone with the Dirac point at
$\left\{k_x, k_z \right\}=\left\{0, \pi \right\}$.

The presence of this single surface Dirac cone can also be seen as  a
consequence of the fact that the bulk three-dimensional Hamiltonian is
characterized by a non-zero mirror Chern number~\cite{teo08} $\mathcal{C}_{\mathcal{M}}=1$
at $k_z=\pi$. When considering systems with a finite number of layers still
preserving reflection symmetry in the stacking direction --  this constraint is
fulfilled for stacks with an odd number of layers -- the surface spectrum in
the  remaining translational invariant $k_x$ direction will therefore display
a single uncoupled chiral mode  as
schematically shown in Fig.~\ref{fig:schematic}(b). We emphasize that this
chiral anomaly is regularized by the presence of a chiral mode partner with
opposite chirality at the opposite $(0{\bar 1}0)$ surface.

We next ``trivialize" our system by introducing a perturbation that explicitly
breaks the mirror symmetry along the stacking direction. To this end, we
consider a dimerization pattern  in the interlayer coupling which further
modifies the effective surface Hamiltonian as ${\cal H}_S={\cal H}_0 +
\left[t_1+t_2 \cos{(k_z)}\right] \sigma_x - t_2 \sin{(k_z)} \sigma_y$ [c.f.
Fig.~\ref{fig:schematic}(a)]. This, in turn, implies that the single surface
Dirac cone acquires a mass $\propto t_1-t_2$ and the surface becomes a
conventional gapped one.   When considering, as before, a finite system with
an odd number of layers there will be a single chiral mode traversing the
full surface gap [c.f. Fig.~\ref{fig:schematic}(c)] localized at one of the
two boundary layers depending upon the specific dimerization pattern. The
existence of this chiral hinge mode  can be understood 
by considering the effective surface Hamiltonian as a collection of one-dimensional Rice-Mele models~\cite{ric82,xia10}, 
parametrized by $k_x$. 
 For a chain
with an odd number of sites, the latter displays an in-gap boundary state at an
energy corresponding precisely to the staggered chemical potential $\equiv
k_x$. If we assume the dimerization pattern to be equivalent at all the four
surfaces perpendicular to the layer planes, the hinge modes at the boundary
layer will be connected  to create a circulating planar current, which is in
agreement with the fact that a dimerized stack of an odd number of layers
defines a (thicker) two-dimensional Chern insulating state~\cite{fu07}.

Let us now instead assume that the dimerization patterns at the two opposite
surfaces $(010)$ and $(0{\bar1}0)$ are designed to be opposite to each other
as shown in Fig.~\ref{fig:schematic}(a). Although still breaking mirror
symmetry along the stacking direction, this configuration preserves bulk
inversion symmetry with the inversion center lying at the center of one layer.
The presence of inversion symmetry also implies that the chiral hinge modes
related to the $(010)$ and $(0{\bar 1}0)$ surfaces will be localized on
opposite boundary layers. The same clearly holds true at the $(100)$ and
$({\bar 1}00)$ surfaces. Moreover, the Jackiw-Rebbi mechanism~\cite{jac76} guarantees the existence
of an additional chiral hinge mode at two inversion- symmetry related hinges
between the $x$ and $y$ planes, and thus the configuration of in-gap hinge states
schematically shown in Fig.~\ref{fig:schematic}(c) is realized. The latter
represents nothing but the hallmark of a second-order topological insulator in
three-dimensions protected by inversion symmetry.

\paragraph{Quantum anomalous Hall stacks -- }   
Having presented our coupled-layer low-energy approach, we next introduce a
microscopic model that features an inversion-symmetric higher-order topological insulating state.
Specifically, we consider stacks of honeycomb
layers, each of which possesses chiral orbital currents leading to a quantum
anomalous Hall (QAH) insulating state~\cite{hal88}. In order to have
alternating Chern numbers on the honeycomb layers, we further assume the
direction of the orbital currents to be opposite in two consecutive layers.
For uncoupled layers, the corresponding tight-binding Hamiltonian reads

\begin{figure}[tbp]
\includegraphics[width=1\columnwidth]{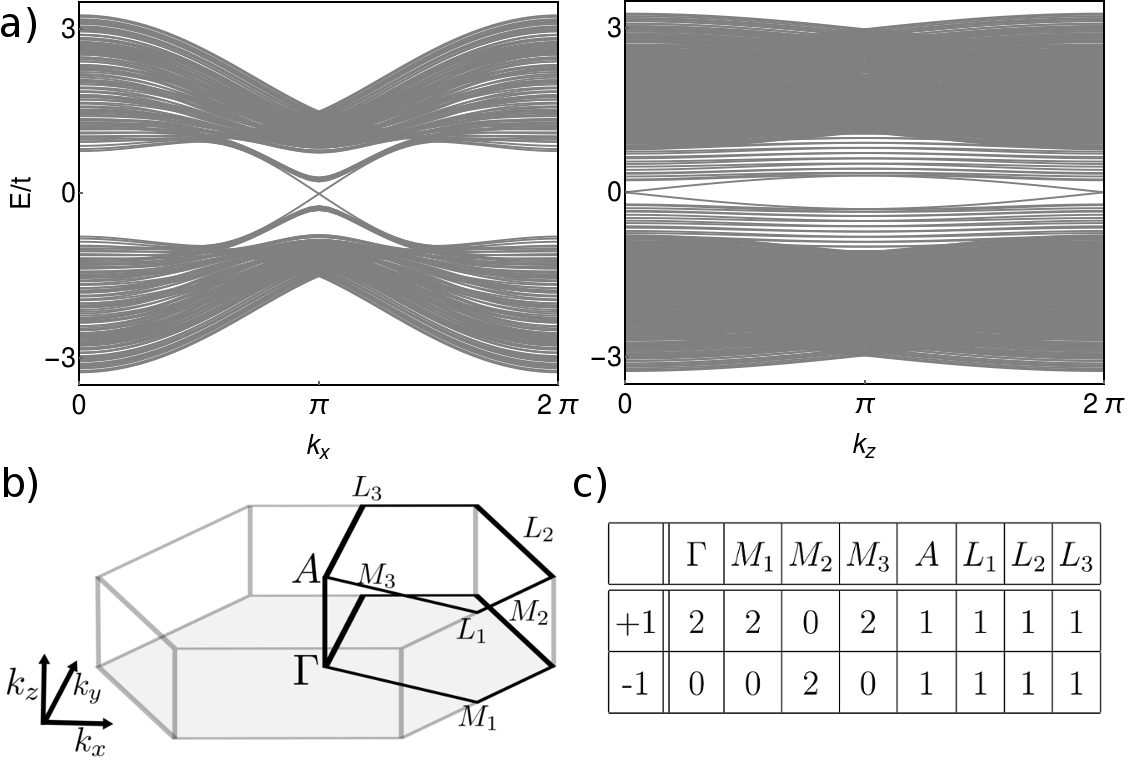}
\caption{a) Edge spectra of the QAH model with periodic boundary conditions in the $x$- and $z$-direction
for stacks of zigzag-terminated honeycomb flakes. The tight-binding Hamiltonian parameters have been chosen as $t_{2}/t=0.2$ and $t_{z}/t=0.3$. b) The hexagonal 3D Brillouin 
zone with the inversion symmetric points labeled. c) Table specifying the number of 
bands with inversion eigenvalue +1 and -1 at the inversion symmetric points
in the Brillouin zone. \label{fig:haldane}}
\end{figure}

\begin{equation*}
{\cal H}_{\parallel}  =-t\sum_{\left\langle i,j\right\rangle ,\alpha}c_{i\alpha}^{\dagger}c_{j\alpha}-it_{2}\sum_{\left\langle \left\langle i,j\right\rangle \right\rangle ,\alpha}(-1)^{\alpha} \nu_{i j} c_{i\alpha}^{\dagger}c_{j\alpha},
\label{eq:haldane}
\end{equation*}
where $c_{i \alpha}^{\dagger}$ ($c_{i \alpha}$) creates (annihilates) an electron on site $i$ in layer $\alpha$,
$t$ is the intralayer nearest-neighbor hopping amplitude and $t_2$ is the
next-nearest neighbor hopping amplitude. As usual, the factor $\nu_{i j} =1$
if the next-nearest neighbor hopping path rotates counterclockwise, and $-1$
if it rotates clockwise.  We next introduce an interlayer coupling that
explicitly breaks the mirror symmetry in the stacking direction but still
preserves bulk inversion symmetry with the inversion center in one layer at
the center of the bond between the two $A$ and $B$ honeycomb sublattices. In
its simplest form the interlayer Hamiltonian is then
\begin{eqnarray}
{\cal H}_{\perp}&=& -\frac{t_{z}}{2}\sum_{i\in A,\alpha} \left[1- \mathrm{e}^{i \pi \alpha} \right] c_{i\alpha}^{\dagger}c_{i\alpha+1}-\frac{t_{z}}{2}\sum_{i\in B,\alpha}  \left[1+ \mathrm{e}^{i \pi \alpha} \right] \nonumber \\ & & \times  c_{i\alpha}^{\dagger}c_{i\alpha+1} + \it{h.c.} 
\label{eq:hinter}
\end{eqnarray}
As discussed below, this interlayer Hamiltonian is naturally realized assuming buckled honeycomb layers. 
Fig.~\ref{fig:haldane}(a) shows the edge energy spectrum as obtained by
diagonalizing the full Hamiltonian ${\cal H}={\cal H}_{\parallel} + {\cal
H}_{\perp}$ for stacks of zigzag terminated ribbons with an odd number of
layers. It agrees perfectly with the foregoing low-energy description. Inside
the gapped bulk energy bands, we clearly observe conventional surface states,
corresponding to a massive surface Dirac cone, in the gap of which two chiral
hinge modes localized on opposite layers appear. Precisely the same features
occur considering periodic boundary conditions in the stacking direction and zigzag terminations in the other two directions. 
 We
point out that we excluded from our analysis ribbons with armchair
terminations since the latter would yield an unprotected single massless
surface Dirac cone. This is due to  the fundamental difference between the chiral edge states of a QAH insulator for zigzag and armchair terminated ribbons~\cite{can13}. For the latter, the chiral edge states have an equal
amplitude on both the two honeycomb sublattices. The interlayer
coupling Hamiltonian introduced above thus yields an effective mirror-symmetric
coupling between the QAH chiral edge states.  However, additional symmetry-
allowed terms in the bulk Hamiltonian, {\it e.g.} intralayer real next-
nearest-neighbor hopping amplitudes, will gap the surface Dirac cone leading
to the observation of chiral hinge modes even for stacks of armchair
terminated ribbons.

To prove the topological origin of these chiral hinge states, we have calculated the bulk $\mathbb{Z}_2$ topological invariant for a second-order topological insulator with inversion symmetry~\cite{mie18}. It can be derived using the bulk formulation of the quantized corner charges for the effective two-dimensional inversion-symmetric Hamiltonians at $k_z=0,\pi$, and thereafter considering the corresponding corner charge flow. When expressing the corner charges in terms of the multiplicities of the inversion symmetry eigenvalues $\pm 1$ of the occupied bands at the inversion-symmetric momenta of the 3D Brillouin zone [c.f. Fig.~\ref{fig:haldane}(b)], we then find the following expression for the bulk $\mathbb{Z}_2$ invariant: 

\begin{align}
\nu & =\left[-\Gamma_{1}-\frac{1}{2}\Gamma_{-1}+\frac{1}{2}\left(M_{1}\right)_{-1}+\frac{1}{2}\left(M_{2}\right)_{-1}+\frac{1}{2}\left(M_{3}\right)_{-1}\right.\nonumber \\
 & \left.+A_{1}+\frac{1}{2}A_{-1}-\left(L_{1}\right)_{-1}-\frac{1}{2}\left(L_{2}\right)_{-1}-\frac{1}{2}\left(L_{3}\right)_{-1}\right]\mathrm{mod}\,2\label{eq:inv}
\end{align}

\begin{figure}[tbp]
\includegraphics[width=1\columnwidth]{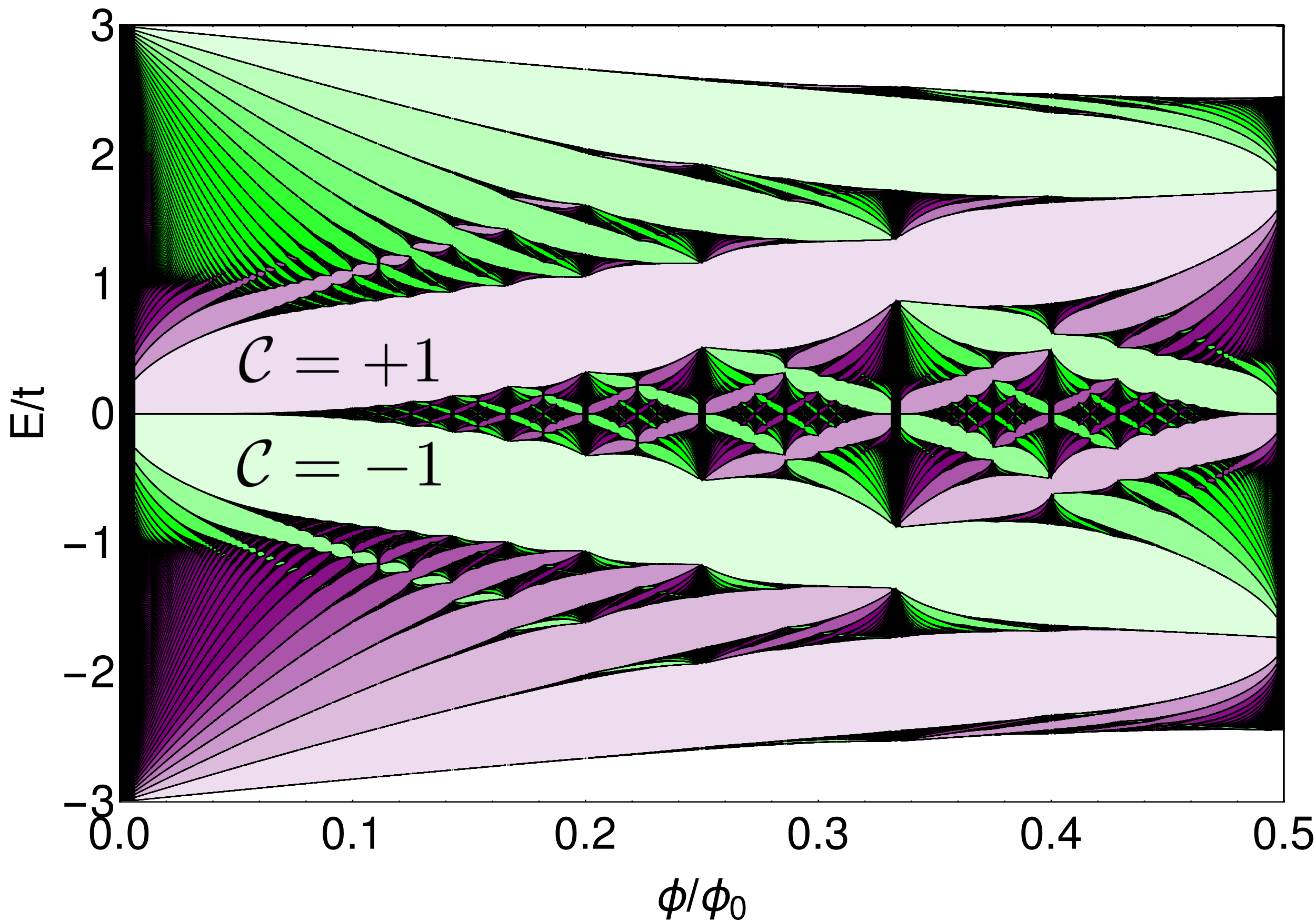}
\caption{Energy versus magnetic flux $\phi$ (measured in units of the magnetic flux quantum $\phi_0$) for the Hofstadter model on a honeycomb lattice. The gaps are coloured according to their Chern number. The two largest gaps have $\mathcal{C}=\pm1$.
 \label{fig:hofstadter}}
\end{figure}

\begin{figure*}[tbp]
\includegraphics[width=2\columnwidth]{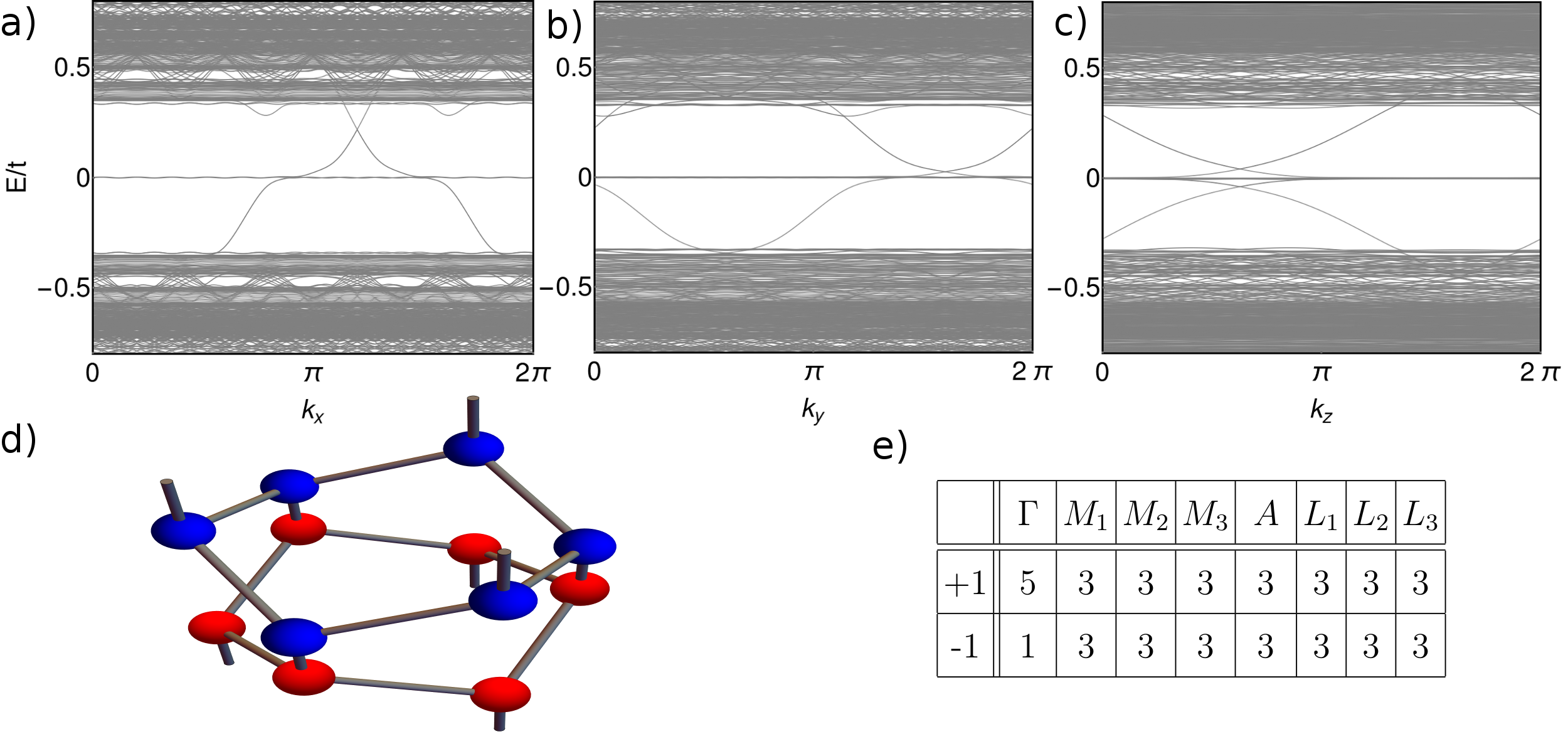}
\caption{a) Spectrum along the zigzag direction, all spectra are for $t_{z}/t=0.5$, $\phi/\phi_{0}=0.2$, $\phi_{2}/\phi_{0}=0.04$, $\phi_{3}/\phi_{0}=0.04$, and $V_{0}=0.6t$,
where $\phi/\phi_{0}$, $\phi/\phi_{1}$ and $\phi/\phi_{2}$ are the flux per plaquette in the $z$, $x$ and $y$ direction respectively.
b) Spectrum along the armchair direction.
c) Spectrum along the stacking direction.
d) Schematic side-view of an AA stacked buckled honeycomb lattice. In the first layer the A sublattice is connected to the next layer, while in the second layer the B sublattice is connected to the next layer. 
e) Table showing the multiplicity of inversion eigenvalues at half-filling at inversion-symmetric points
in the Brillouin zone for $\phi/\phi_{0}=1/3$, $V_{0}=0.8t$. \label{fig:qhe2}}
\end{figure*}

A non-trivial value $\nu=1~\textrm{mod}~2 $ of this invariant guarantees the presence of chiral hinge modes provided the bulk and the surfaces are gapped. 
For our model at half-filling, the inversion symmetry labels  [c.f. Fig.~\ref{fig:haldane}(c)] directly imply an higher-order topology. Therefore the chiral hinge states shown in Fig.~\ref{fig:haldane}(a) are the direct manifestation of a bulk-hinge correspondence.

\paragraph{Stacks of quantum Hall layers --} With these results in hand, we next introduce our main result: the possibility to engineer an inversion symmetry protected second-order topological insulator in stacks of doped quantum Hall layers with a buckled honeycomb geometry. To show this, we first recall that the quantum Hall effect on the honeycomb lattice exhibits a zeroth Landau level above (below) which the total Chern number $\mathcal{C}=+1(-1)$ for relatively weak magnetic fluxes per plaquette [see
Fig.~\ref{fig:hofstadter}]. This also implies that the sign of the Hall conductance is opposite in p- and n-doped layers. We then take advantage of this property to realize a quantum Hall analogue of the quantum anomalous Hall model introduced above. For the intralayer part of the Hamiltonian we thus consider layers with an alternating bias $\pm V_0$. The corresponding Hamiltonian is then 
\[
\mathcal{H}_{\parallel}=-t\sum_{\left\langle i,j\right\rangle ,\alpha}c_{i,\alpha}^{\dagger}c_{j,\alpha}+V_{0}\sum_{i,\alpha}\left(-1\right)^{\alpha}c_{i,\alpha}^{\dagger}c_{i,\alpha}.
\]
The effect of the perpendicular magnetic field is taken into account via the usual Peierls substitution $t\rightarrow t\,e^{i\int
d\mathbf{s}\,\mathbf{A}},$ where $d\mathbf{s}$ is the line integral between
the bonds and we took the vector potential in the Landau gauge
$\mathbf{A}=-(eB/\hbar)(y,0,0)$ with $e$ the electron charge and $B$ the
magnetic field strength. The interlayer Hamiltonian is taken to be the same as in Eq.~\eqref{eq:hinter}
to account for the buckling of the layers [c.f.  Fig.~\ref{fig:qhe2}d)].

We have first verified the non-trivial topology of this microscopic tight-binding model by computing the multiplicities of the inversion symmetry
eigenvalues at the inversion-symmetric momenta of the BZ [see
Fig.~\ref{fig:qhe2}(e)]. When computing Eq.~\eqref{eq:inv} we then find a non-
trivial value of the $\mathbb{Z}_2$ topological invariant. However, this model
Hamiltonian does not possess gapped surfaces. This is a consequence of the
fact that the honeycomb sublattice symmetry is not broken by the external
magnetic field. Therefore, an effective coupling between the quantum Hall edge
states is absent. Clearly, sublattice symmetry breaking terms, the simplest of
which are real intralayer next-nearest neighbor hoppings [see the Supplemental
Material], effectively open a scattering channel between the intralayer edge
modes and thus will yield conventional gapped surface Dirac cones.

An alternative approach relies on tilting the direction of the external
magnetic field away from the stacking direction towards the $[111]$ direction.
As evident from Fig.~\ref{fig:qhe2}(a-c) such a magnetic field direction
tilting allows to open substantial surface gaps. Moreover, the chiral hinge states at positive and negative energies are separated by a zeroth surface Landau level, and therefore live on different hinges [see the Supplemental Material]. As discussed
above, this change in the location of the chiral hinge modes by tuning the Fermi level is perfectly compatible with the higher-order topology of our system.

\paragraph{Conclusion --} To sum up, we have proved the existence of a second-order topological
insulator protected by inversion symmetry using a coupled layer approach in
which the layers have alternating Chern numbers. It can be derived
from a parent topological mirror Chern insulator by crystalline symmetry
breaking terms that retain the bulk inversion symmetry of the bulk crystal.
The presence of the topologically protected chiral hinge modes can be
inferred from a three-dimensional ${\mathbb Z}_2$ invariant. We have shown
that a non-trivial value of this invariant can be encountered in stacks of
doped quantum Hall layers with a buckled honeycomb lattice structure.
As a result, we believe that silicene multilayers provide an excellent
platform to engineer chiral inversion-symmetric higher-order topological insulators.

\paragraph{Acknowledgements --} C.O. acknowledges support from a VIDI grant (Project 680-47-543) financed by the Netherlands Organization for Scientific Research (NWO). This work is part of the research programme of the Foundation for Fundamental Research on Matter (FOM), which is part of the Netherlands Organization for Scientific Research (NWO). S.K. acknowledges support from a NWO-Graduate Program grant.

\newpage
\onecolumngrid

\section*{Supplemental Material}

In this supplemental material we show that a real nearest-neighbor hopping in
one of the buckled honeycomb layers is enough to open a surface gap in the model of doped quantum Hall
layers presented in the main text. In addition, we briefly discuss the
localization of the hinge states present.

\subsection*{Surface gap opening by nearest-neighbor hopping}

Let us consider one layer of silicene in a zigzag ribbon geometry
with a perpendicular magnetic field threading a flux $\phi$, at a
bias $V_{0}$ and with real next-nearest neighbor hopping $t'$. The
Hamiltonian is 

\begin{align}
H(k_{x}) & =\sum_{j}\psi_{k_x,j}^{\dagger}\begin{pmatrix}V_{0}-2t'\cos\left(k_{x}\right) & -t'\left(e^{i\pi\phi j}+e^{ik_{x}}e^{-i\pi\phi j}\right)\\
-t'\left(e^{-i\pi\phi j}+e^{-ik_{x}}e^{i\pi\phi j}\right) & V_{0}-2t'\cos\left(k_{x}\right)
\end{pmatrix}\psi_{k_x,j}\nonumber \\
 & -\sum_{j}\psi_{k_x,j+1}^{\dagger}\begin{pmatrix}t' & t\\
0 & t'
\end{pmatrix}\psi_{k_x,j}+h.c,\label{eq:}
\end{align}
where $\psi_{k_x,j}=\left(a_{k_x,j},b_{k_x,j}\right)$, with $a_{k_x,j}$ and $b_{k_x,j}$
the annihilation operators on the A and B site of unit cell $j$ respectively.

If we consider a stack of such layers with alternating $\pm V_{0}$, we
will have a gapped surface if the edge state of a layer hybridizes
differently with the edge state of the layer above and below it. This
corresponds to $t_{1}\neq t_{2}$ in the effective surface model presented
in the main text. To check whether this is the case we take the Hamiltonian
Eq.~(\ref{eq:}), and fix $V_{0}$ and $\phi$ such that the Fermi
energy lies in the gap with Chern number $+1$ (see Fig.~3 in the
main text). We solve for the right-moving edge state at $E=0$. We
then also take the Hamiltonian at $-V_{0}$ and solve for the left-moving
edge state at $E=0$ at the same momentum $k_{x}$. Let us denote
these states by $\left|\psi\right\rangle $ and $\left|\chi\right\rangle $.
There are two different hoppings between two layers: one connecting
all the A sublattice sites, and one connecting all the B sublattice
sites. Let us denote these by $H_{\perp A}$ and $H_{\perp B}$. If
the mixing due to these is different, 
\begin{align}
\left|\left\langle \chi\right|H_{\perp A}\left|\psi\right\rangle \right| & \neq\left|\left\langle \chi\right|H_{\perp B}\left|\psi\right\rangle \right|,\label{eq:-1}
\end{align}
there is an effective dimerization and a gapped surface.

If $t'=0$, $\left|\psi\right\rangle $ and $\left|\chi\right\rangle $
are related to each other by chiral symmetry. In addition we have
sublattice symmetry, which means that the edge states have equal weight
on the A and B sublattices. From this it follows that Eq.~(\ref{eq:-1})
is not satisfied. Taking $t'\neq0$ breaks sublattice symmetry and
leads to edge states that do not have equal weight on the A and B
sublattices. We now take even layers to have $t'/t=0.1$, $V_{0}=-0.8 t$,
$\phi/\phi_0=1/5$ and the odd layers to have $t'=0,$ $V_{0}=0.6 t$, $\phi/\phi_0=1/5$ (since $t'\neq0$
breaks particle-hole symmetry we slightly adjust $V_{0}$ to remain
in the gap). Solving for $\left|\psi\right\rangle $ and $\left|\chi\right\rangle$ then gives

\begin{align*}
\frac{\left|t_{1}\right|}{\left|t_{2}\right|}=\frac{\left|\left\langle \chi\right|H_{\perp A}\left|\psi\right\rangle \right|}{\left|\left\langle \chi\right|H_{\perp B}\left|\psi\right\rangle \right|}\approx0.985 & .
\end{align*}
This means that the surface is gapped. We note that the mixing is
small, but it nevertheless shows that the gapless surfaces are not
protected and can be gapped out by symmetry-allowed perturbations.

\newpage

\subsection*{Localization of hinge states}

\begin{figure}[h]
\includegraphics[width=1\columnwidth]{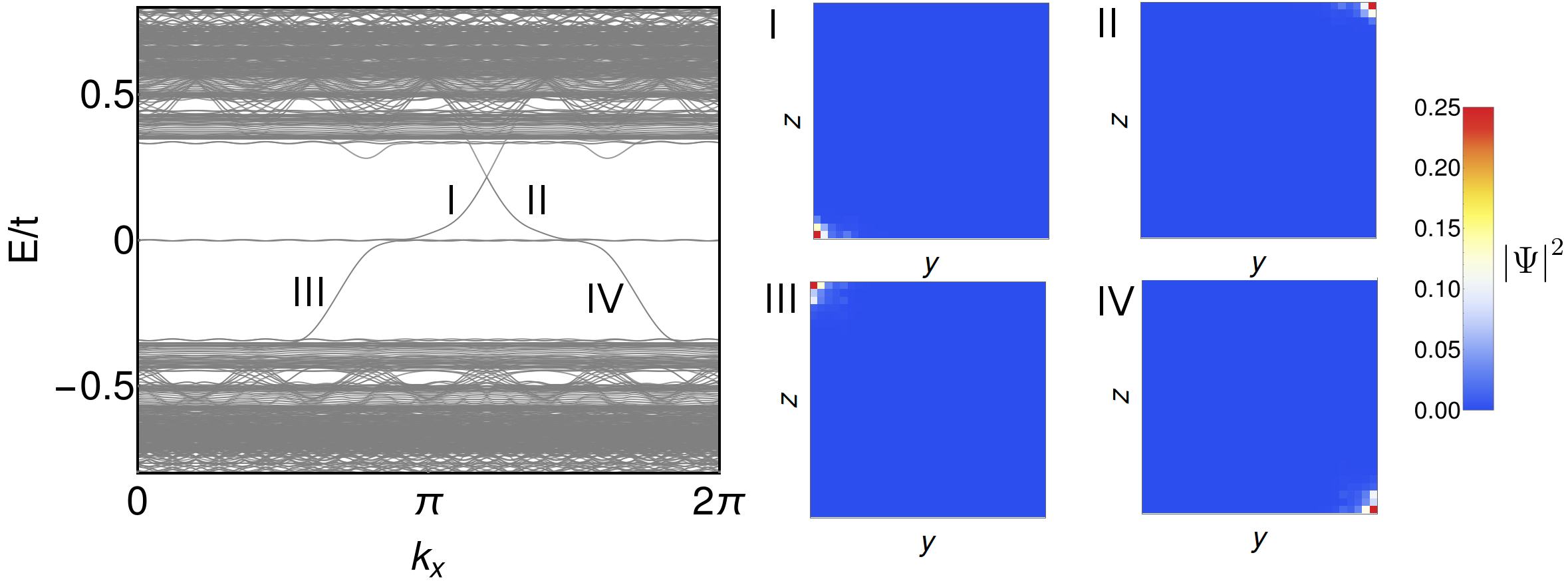}
\caption{Figure showing the localization of the hinge states along the armchair direction. Going from the gap above the zeroth LL of the surface to the one below it, the chiral hinge states localize on different hinges. This realizes both inversion symmetric configurations, as alluded to in the main text. \label{fig:localization}}
\end{figure}

\end{document}